\documentclass{emulateapj}

\usepackage{graphicx}
\newcommand{\Kepler}{\emph{Kepler\ }}

\usepackage{commath}
\usepackage{array}

\newcolumntype{C}[1]{>{\centering\let\newline\\\arraybackslash\hspace{0pt}}m{#1}}

\begin{document}

\title{Mass, Density, and Formation Constraints in the Compact, Sub-Earth Kepler-444 System including Two Mars-Mass Planets}
\author{Sean M. Mills and Daniel C. Fabrycky}
\email{sean.martin.mills@gmail.com}
\affil{The Department of Astronomy and Astrophysics\\ The University of Chicago\\ 5640 S. Ellis Ave, Chicago, IL 60637, USA}

\begin{abstract}

Kepler-444 is a five planet system around a host-star approximately 11 billion years old. The five transiting planets all have sub-Earth radii and are in a compact configuration with orbital periods between 3 and 10 days. Here we present a transit-timing analysis of the system using the full \Kepler data set in order to determine the masses of the planets. Two planets, Kepler-444 d ($M_\mathrm{d}=0.036^{+0.065}_{-0.020}M_\oplus$) and Kepler-444 e ($M_\mathrm{e}=0.034^{+0.059}_{-0.019}M_\oplus $), have confidently detected masses due to their proximity to resonance which creates transit timing variations. The mass ratio of these planets combined with the magnitude of possible star-planet tidal effects suggests that smooth disk migration over a significant distance is unlikely to have brought the system to its currently observed orbital architecture without significant post-formation perturbations.

\end{abstract}

\section{Introduction}

Probing the mass-radius relationship for planets smaller than Earth is interesting to theorists as it may be used to constrain the formation and composition of these bodies, a topic of debate in the current literature \citep[e.g][]{2010apf..book.....A,2010exop.book..297C,2013SSRv..180...71S,2014ApJ...780...53C,2016ApJ...817...80D}. A few planets in this size regime have been characterized \citep[e.g.][]{2013SSRv..180...71S,2013ApJ...773L..15R,2015Natur.522..321J,2017Natur.542..456G}; however, due to the small number of characterizable systems, little is yet known about the masses or compositions of the smallest ($\lesssim1R_\oplus$, $\lesssim1M_\oplus$) planets, despite them being among the most common in the galaxy \citep{2015ApJ...808...71M}.

Recent work has demonstrated the effectiveness of using photodynamic modeling to extract transit timing variations (TTVs) and planetary properties from systems with a low signal-to-noise ratio (SNR) \citep[e.g.,][]{2012Sci...337..556C,2015MNRAS.454.4267B,2016Natur.533..509M}. This technique takes advantage of the many transits of short-period planets observed in the \Kepler data by fitting the entire light curve and all transits simultaneously. Here we apply this technique to Kepler-444. 

Kepler-444's planets (b, c, d, e, and f from inside to out) range in radii from 0.4 to 0.8 $R_\oplus$ and in orbital period from 3.6 to 9.8 days \citep{2015ApJS..217...16R,2015ApJ...799..170C}. Their period ratios are near, but not exactly on, mean motion resonances (MMRs; see Table~\ref{table:prats}). Despite the compact architecture of the system, it is around a star $11.2\pm1.0$ Gyr old \citep{2015ApJ...799..170C} and therefore has likely been in a stable configuration for billions of years. A tight binary pair of M-dwarf stars also orbit together around Kepler-444 with a period of approximately 460 years and a distance of $\sim$ 60 AU \citep{2015ApJ...799..170C}. Such a configuration poses a puzzle regarding the early history of the Kepler-444 system, as planetary formation and migration in a truncated protoplanetary disk in the presence of a very nearby binary star pair is not well understood, with several effects newly proposed \citep[e.g.,][]{2015Natur.524..439T,2016MNRAS.459.2925X}. Recent studies have attempted to understand the possible histories of the system and use it to place constraints on formation mechanisms \citep{2016ApJ...817...80D,2016CeMDA.tmp...21P}. However, such studies were unable to use the actual compositions or masses of the Kepler-444 planets since they were hitherto unknown. In this paper, we use photodynamics to put constraints on the masses of the planets in the Kepler-444 system and report mass detections for two of the planets: $M_d=0.036^{+0.065}_{-0.020}M_\oplus$ and $M_e=0.034^{+0.059}_{-0.019}M_\oplus$.

\section{Methods}

\label{sec:methods}

\begin{figure}
\centerline{
\includegraphics[scale=0.7]{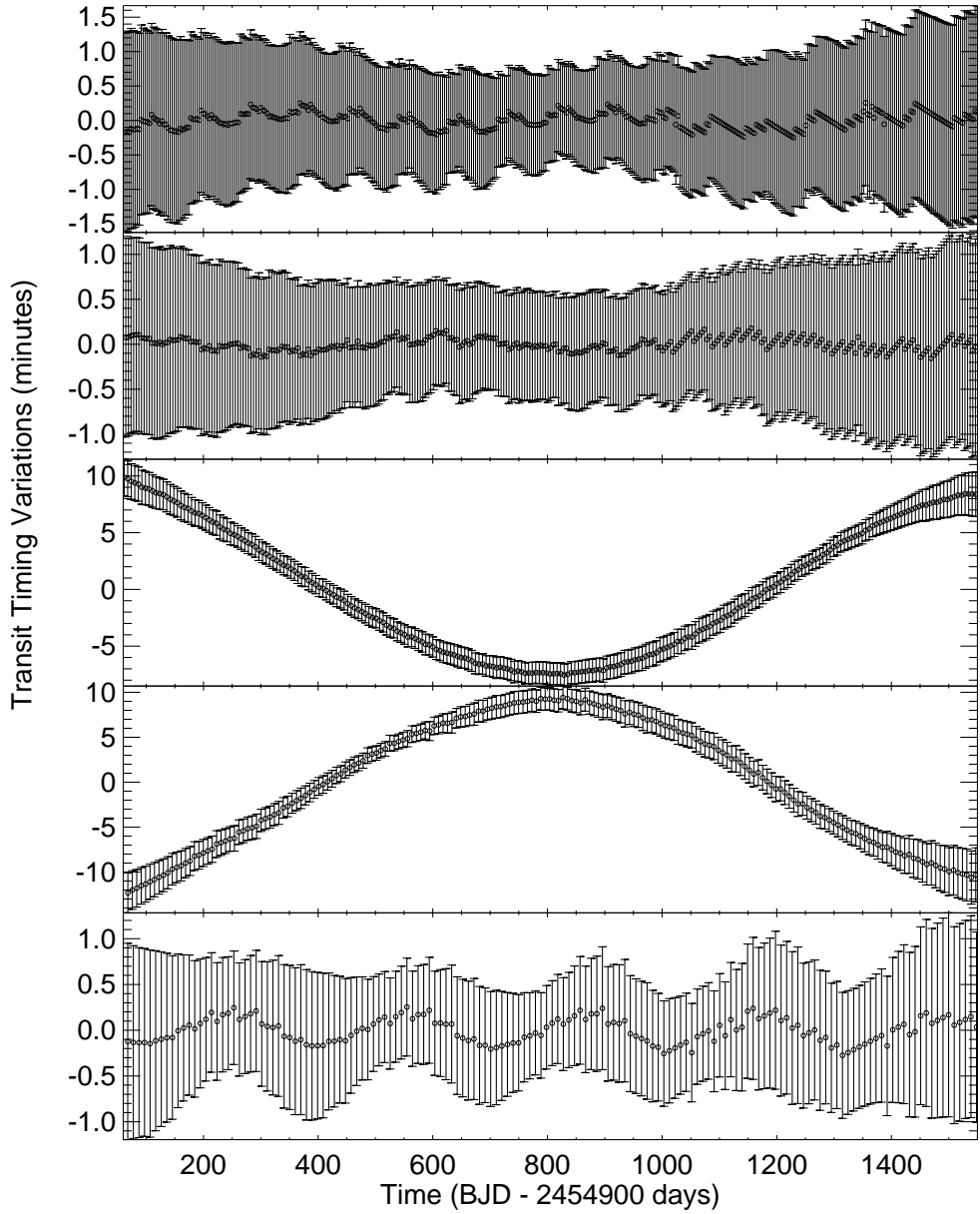}
}
\caption{
TTVs and uncertainties as measured by the photodynamic DEMCMC as described in \S\ref{sec:methods}. The values and error bars were generating by drawing from 100 parameter sets from the posterior and integrating the equations of motion to generate median and 1-$\sigma$ uncertainties. %The gray curves are best fit sin curves, corresponding to the dominant TTV signals from neighboring planets. 
The anti-correlated TTV signal between planets d and e with a $\sim$10 minute amplitude is readily visible by eye and results in a secure mass detection for both planets. All other TTV signals are below the noise level. 
 }
\label{fig:photottv}
\end{figure}

We initially identified potential transit timing variations in the Kepler-444 system by simultaneously fitting the raw \Kepler light curve with a planet transit model \citep{2002ApJ...580L.171M} and a 1-day wide polynomial to take into account systematic effects and stellar activity. %We plot the results in Fig.~\ref{fig:danttv} and note that we see anti-correlated TTVs between planets d and e. 
We also compute the expected period of the TTV signal between each pair of planets analytically \citep{2012ApJ...761..122L} (Table~\ref{table:prats}), noting that period of the expected signal for planets d and e matches the TTV observations well (Fig.~\ref{fig:photottv}). %The TTV period between the other pairs of neighboring planets is much smaller than the length of the observing window due to the pairs' greater distance from resonance and therefore potentially observable; however, their amplitude is predicted to be undetectably small. Indeed 
We find statistically significant TTV between planets d and e, but the signal for all other planets is undetectably low as theoretically expected. The same conclusion was reached independently by \cite{2016arXiv161103516H}, a survey of many \Kepler systems showing TTVs.

In order to perform a more robust, simultaneous fit for all planetary parameters, we first reprocessed the raw \Kepler lightcurve data. We use short-cadence (58.8 second integration) data when it was available (\Kepler observing quarters 4, 6, and 15-17) and long cadence data (29.4 minute integrations) otherwise. We first discarded points whose quality flag had a value equal to or greater than 16. We then detrended the light curves by masking out the expected transit times plus 20\% of the transit duration to account for possible TTVs and then fit a cubic polynomial model with a 1000-minute width centered on photometric data points spaced by 30 minute intervals. We interpolated between these points to determine a baseline and divide the measured flux at each data point by these values. This detrending method produces two regions of extreme curvature in the lightcurve due to edge effects, so we discard the small regions with times BJD-2454900 = 1405.10 to 1405.18 days and 1490.88 to 1490.97 days. To account for certain Quarters showing higher noise levels than others despite all quarters having similar quoted uncertainties, we assign an uncertainty of 5.3030402e-05 to points in Quarter 12, 2.3470900e-04 in Quarter 16, and 6.5361999e-04 in Quarter 17, an increase over other regions by a factors of roughly 5, 4, and 11 respectively (the ratio of their out of transit standard deviation). Lastly, we increase the uncertainties in all points by a multiplicative factor of 1.38073 so that a fiducial fit to the light curve has a $\chi^2=1.000$. This approach substantially increases the uncertainties on the fitted parameters compared to using the values reported by \Kepler, allowing unmodeled noise to propagate to the final uncertainties on our parameter posteriors. We used data from \Kepler Data Release 21 (DR21) for this analysis because there is less scatter in the DR21 data compared to the DR25.

Our photodynamic model integrates Newtonian equations of motions for the star and five planets including the light travel time effect (which in this case is negligible). When any of the planets pass in front of the star along the line of sight, a synthetic light curve is generated \citep{2012MNRAS.420.1630P}, which can then be compared to the data. The parameters we include for each planet in the differential evolution Markov chain Monte Carlo \cite[DEMCMC;][]{TerBraak2005} fit are $\{P,T_0,e^{1/2}\cos(\omega),e^{1/2}\sin(\omega),i,\Omega,R_p/R_\star,M_p/M_\star\}$, where $P$ is the period, $T_0$ is the mid-transit time, $e$ is eccentricity, $\omega$ is the argument of periapse, $i$ is inclination to the sky plane, $\Omega$ is nodal angle on that plane, and $R$ and $M$ are radius and mass respectively (with subscripts $p = b,c,d,e,f$ for the planets and $\star$ for the star). The star had five additional parameters: $\{M_\star,R_\star,c_1,c_2,dilute\}$, where $c_i$ are the two quadratic limb-darkening coefficients and $dilute$ is the amount of dilution from other nearby sources. 

We put physically sensible, but permissive, minima ($\rho_p=0.0$) and maxima ($\rho_p=\rho_{\mathrm{Fe}}$) on the bulk planet densities, where $\rho_p$ is a planet's bulk density and $ \rho_{\mathrm{Fe}}$ is the density of iron for a body of planet p's size. Taking values from \cite{2007ApJ...669.1279S}, the maximum densities for the 5 planets from b to f respectively are (9.5, 9.7, 10.3, 10.5, 12.2) g/cm$^3$, differing due to the compressibility of iron. The prior on mass is otherwise flat between 0 and these values.%, which is a broad prior given that we do not expect arbitrarily low density planets. 

Since the mass and eccentricity implied by TTVs may be degenerate \citep{2012ApJ...761..122L} and result in measured eccentricity values so high that the system go unstable on timescales much shorter than the age of the system \cite{2015ApJ...807...44P}, we use a Rayleigh prior on the eccentricity of all planets with width parameter $\sigma=0.02$. This is consistent with the values measured in other tightly packed planetary systems \citep{2014ApJ...787...80H,2014ApJ...790..146F}, and is consistent with long term stability because even moderate eccentricity has been shown to destabilize tightly packed systems \citep{2015ApJ...807...44P}.

The dilution is well-measured \citep{2015ApJ...799..170C} so we fix $dilute= 0.0394$ since it would otherwise be highly degenerate with $R_p/R_\star$. This implies that the $R_p/R_\star$ value uncertainties may be slightly underestimated, but since \cite{2015ApJ...799..170C} report $dilute=0.0394 \pm 0.0001$, this will have only a very small effect on the reported posteriors. We also fix $\Omega_p=0$ for all planets since we expect very small mutual inclinations between the planets because we see five planets transit \cite[see, e.g.,][]{2011Natur.470...53L}. Additionally, even modest values of $\Omega$ may greatly increase likelihood that the system becomes unstable over the system's lifetime due exchange of eccentricity and inclination on secular time scales. Since the transit information gives only the stellar density and planet-to-star mass ratio (via TTVs), we model with a fixed $M_\star=0.758M_\odot$, which sets the overall scale of the system. We use generic flat priors in all other parameters.
%Since the planet mass uncertainties dominate the stellar uncertainties in this case, this is a reasonable approximations, and all reported parameters and uncertainties may be scaled with current and future stellar measurements.

\section{Results}

\begin{figure}
\centerline{
\includegraphics[scale=0.6]{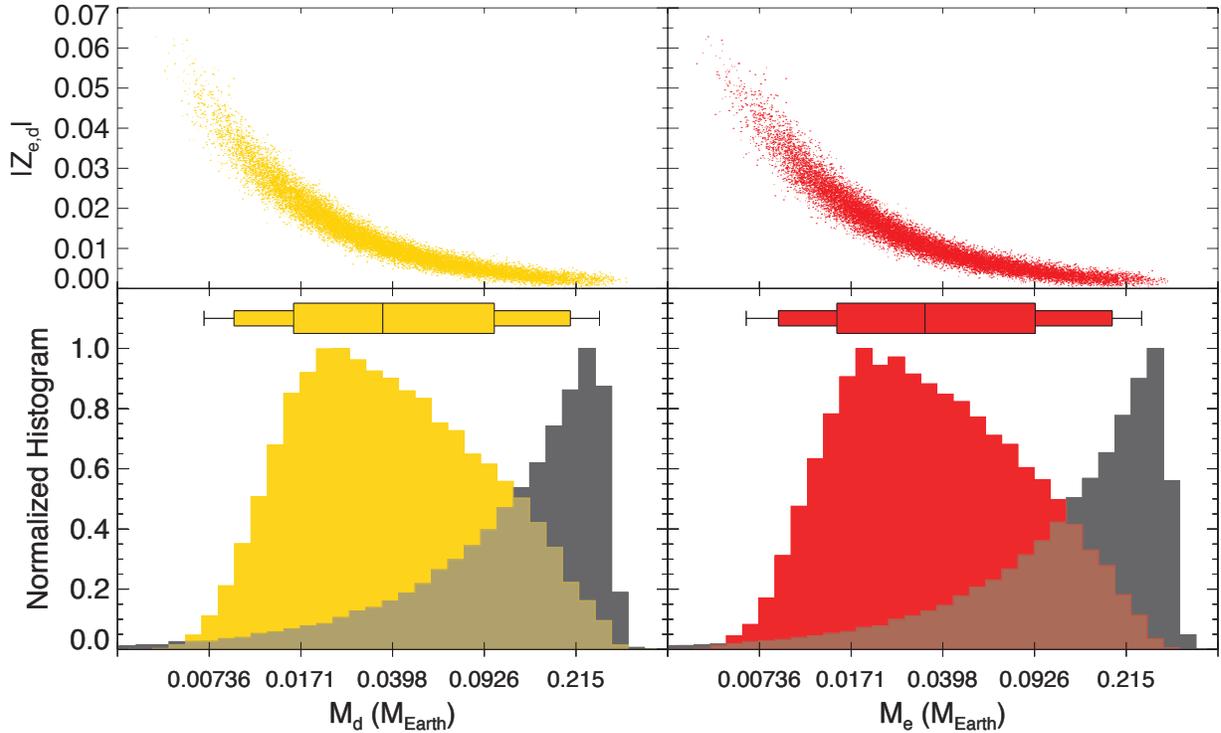}
}
\caption{
The posterior distributions of the masses of planets d (left, yellow) and e (right, red) against the $Z_{e,d}$ posterior (top) and marginalized over all parameters (bottom). The distribution of the mass prior for each planet is plotted in gray (note that the prior is flat in linear space). Box-and-whisker figures show the median, 68.3\%, 95\%, and 99\% confidence intervals above the bottom panels. These panels illustrate how the posteriors cut off more rapidly than the prior at very low masses and also disfavor large masses because of the declining probability on the right hand of the distributions despite the increasing prior.
}
\label{fig:mass}
\end{figure}

Median values and 68.3\% confidence intervals from the of photodynamic model are reported in Table 2. The full data set of the DEMCMC chains can be downloaded from the online version of this article. We ran a 64-chain DEMCMC for 900,000 generations recording every 1,000 generation, conservatively throwing out the first 50,000 generations as a burn-in. The autocorrelation timescale for the slowest converging parameters was approximately 60,000 generations, thus we are left with $\gtrsim 850$ independent samples for each parameter. By numerically fitting the TTVs, the model produces mass constraints based on the \Kepler data. Consistent with the measurement of individual transit times described in \S\ref{sec:methods}, planets b, c, and f do not induce significant TTVs on the other planets, which means their masses are not significantly detected. However, the TTVs in planets d and e are both significant enough to confidently place upper and lower bounds on the mass. The posteriors in mass are inconsistent with zero mass and fall off much more rapidly than the prior near $m=0$. Photodynamically measured TTVs are shown in Fig.~\ref{fig:photottv}, with the resulting mass constraints for planets d and e shown in Fig.~\ref{fig:mass} and reported for all planets in Table~\ref{table:allparams}. Compared to the masses derived in \citet[][$M_d=0.2^{+0.5}_{-0.1}M_\mathrm{\oplus}$ and $M_e=0.1^{+0.2}_{-0.1}M_\mathrm{\oplus}$]{2016arXiv161103516H}, these new measurements ($M_\mathrm{d}=0.036^{+0.065}_{-0.020}M_\oplus$ and $M_\mathrm{e}=0.034^{+0.059}_{-0.019}M_\oplus$) are more precise. This is due in part to more a more restrictive and physical prior and in part to due to the photodynamic analysis method used in this study. 

We compute the posterior of $Z_{j+1,j}$ for each neighboring planet pair by approximating the value as $|z_{j+1}-z_j| / \sqrt{2}$ \cite[see, e.g., ][Eq. 4]{2016arXiv161103516H}, where $z_j = e_j e^{i \omega_j}$ for each planet $j$ and $i$ is the imaginary unit. We find the median and 68\% confidence intervals or upper limits $Z_{c,b}=0.022^{+0.013}_{-0.011}$, $Z_{d,c}=0.021^{+0.013}_{-0.010}$, $Z_{e,d}\le0.023$, $Z_{f,e}\le0.020$. We note that the interior planets are consistent with the prior alone, but the planet pair with detected masses (d and e) has a smaller value preferring low free eccentricity (Fig.~\ref{fig:mass}). The absence of measurable TTVs induced by planet f on planet e also constrains $Z_{f,e}$. %For completeness, we show the transits with TTVs removed in Fig.~\ref{fig:transits}.

We also numerically integrate 100 draws from the DEMCMC posterior for 100 Myr to make sure we are exploring regions of parameter space stable for times comparable to a reasonable fraction of the system's age. 95\% of the samples remain stable. Since approximately equal numbers of systems are likely to go unstable in logarithmic bins of time \citep{2015ApJ...807...44P}, we expect $>80$\% of our posterior to be stable for the measured system age of $\sim$11 Gyr.

\begin{figure}
\centerline{
\includegraphics[scale=0.5]{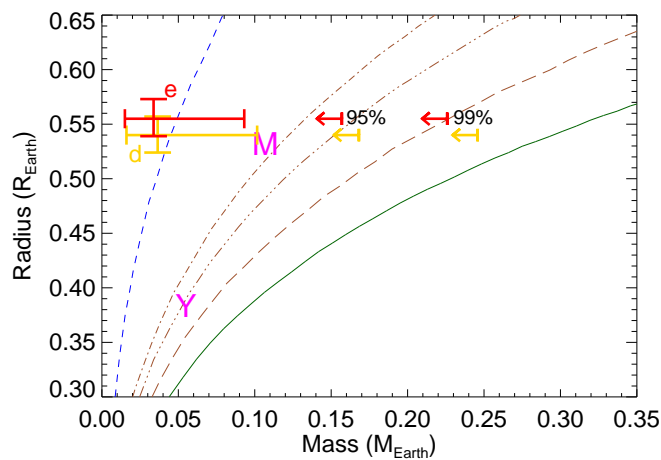}
}
\caption{
The mass and radius (and 1-$\sigma$ uncertainties) of planets d (yellow) and e (red) are plotted on top of contours of constant composition taken from \cite{2007ApJ...669.1279S}. The solid green line represents pure iron planets, the brown lines pure rock (MgSiO$_3$), an Earth-like rock/iron ratio, and a Mercury-like rock/iron ratio from top to bottom, and the blue line represents a pure water planet. The vertical lines with arrows are the 95\% and 99\% upper bounds from the MCMC posterior, showing that the planets are inconsistent with a pure iron composition, and instead have a rockier composition consistent with the Solar System terrestrial planets. Mercury (Y) and Mars (M) are shown in pink. %The black cross is an estimate of possible PLATO mass measurement uncertainties for planets d and e, but offset from their true values for visual clarity. 
}
\label{fig:comp}
\end{figure}

Importantly, we note that at the 95\% confidence level, both planets are inconsistent with being purely iron. Planet d requires a composition with a fraction of rock as least as great as Mercury ($\sim$30\%), and, like Earth, planet e can be no more than 30\% iron by mass. These measurements are plotted in Fig.~\ref{fig:comp} along with theoretical composition tracks taken from \cite{2007ApJ...669.1279S}.

\section{Followup Observations}

\subsection{Radial Velocities}
\label{sec:rvs}

The radial velocity (RV) signal induced on a host star by a planet is given by \cite{1999ApJ...526..890C}:
\begin{equation}
K = \bigg( \frac{2 \pi G}{P} (M_\star+M_\mathrm{p}) \bigg)^{1/3} \frac{M_\mathrm{p}}{(M_\star+M_\mathrm{p})}  \frac{\sin(i)}{\sqrt{1-e^2}}, 
\end{equation} 
where $K$ is the RV amplitude, $G$ the Newtonian gravitational constant, $P$ the planet's period, $M_\star$ the stellar mass, and $M_\mathrm{p}$ the planet's mass. Inserting values for for Kepler-444 planets, we see that the expected $K$ values range from $\sim$4-20 cm/s. This is below the current RV detection threshold  \citep[e.g.,][]{2015arXiv150301770P}.% For instance, \cite{2015arXiv150301770P} show that for a spectrograph with no instrumental noise floor, a 20-minute integration with a 10-meter telescope on a star $m_\mathrm{V}=9$ (Kepler-444  has $m_\mathrm{V} = 8.969$) will have a precision of 20 cm/s, however in practice it has proven challenging to reach these precision levels due to astrophysical and instrumental systematics. 

\subsection{PLATO}
Because of the shallow transit depth, photometric follow-up is precluded for most existing instruments. However, the ESA's Planetary Transits and Oscillations of Stars Mission (PLATO) has recently received approval with operational dates of 2024-2020\footnote{See PLATO SCIRD - http://sci.esa.int/plato/42730-scird-for-plato/.}. The precision goal for PLATO is $3.4\times10^{-5}$ in 1 hour for stars with $m_\mathrm{V}\le11$. Since Kepler-444 is 2 magnitudes brighter, we may expect a factor of $\sim$10 times more photons and thus a precision of $1\times10^{-5}$ per hour. Each planned 50 second exposure should therefore have a precision of $\sqrt{3600/50} \times10^{-5}\approx8\times10^{-5}$. Taking several solutions from the \Kepler data posteriors, based on the planned observing strategy we produce 2-year sets of simulated PLATO transits beginning in 2025. We then add Gaussian noise to this data with $\sigma=8\times 10^{-5}$.
Finally, we refit the combined actual \Kepler data and simulated, noisy PLATO data to test how informative the PLATO measurements will be in further constraining the planet masses. We find that the mass constraints of planets d and e are improved to having $\sim$20\% 1-$\sigma$ uncertainties. Such a measurement may allow tight constraints on the fraction of the planet which is iron, rocky, or volatile, potentially distinguishing a water-rich planet from an Earth-like composition. Additionally, we find that in some cases Planet b (the smallest radius planet $R_b=0.406\pm0.013 R_\oplus$) interacts with Planet c sufficiently to induce observable TTVs and a 99.7\% confidence (3-$\sigma$ equivalents) non-zero mass detection of Planet b. Such a measurement would make it (as of right now) the smallest exoplanet with a detected mass orbiting a main sequence star. To conclude, we note that the results in this section are dependent on the true noise properties and observing strategy of PLATO, which are currently uncertain.

\section{Implications for Formation and Tidal Evolution}
\citet[][hereafter P16]{2016CeMDA.tmp...21P} performs an in-depth analysis of the possible migration history of the Kepler-444 system, considering both migration and circularization effects due to planet-disk interactions. Since the planets are very low mass, P16  assumes they are in the Type I migration regime with migration timescale, $\tau_\mathrm{mig}\propto M_\mathrm{p}^{-1}$. If the planets migrate at different rates (due to mass and local disk density), then one would expect them to approach MMRs with other planets, at which point they would get trapped near those MMRs \citep{1996MNRAS.280..854M,2002ApJ...567..596L,2007ApJ...654.1110T}. Since the planets are up to 2\% away from resonance, P16 speculate that significant relative contraction of the planets did not occur, although significant migration as a unit might have. In order to match the observed period ratios, P16 assumes that planet e is significantly (by a factor of $\sim$3) more massive than d. This allows e to easily migrate more quickly than, and thus contract and approach resonance with, planet d while the other planets remain relatively more distant from resonances. Our photodynamical fit finds that $M_\mathrm{e}/M_\mathrm{d}=0.93^{+0.14}_{-0.13}$, a significant departure from that assumption. 
%We infer from this mass ratio that it is unlikely the planets of this particular system formed from a purely disk migration pathway, but rather either formed in situ or underwent significant orbital period changes after formation. 
This suggests that the present-day observed period ratios combined with smooth disk migration alone are generally insufficient for modeling specifics of the formation of the system. %%
Many factors may have changed the migration of the planets while the disk was present, including local disk properties \citep{2014AA...569A..56C} or turbulence in the disk \citep{2007ApJ...670..805O,2009AA...497..595R}. Alternately, the planets may have moved after the dispersal of the gas and dust disk, for instance via a combination of planetesimal crossings \citep{1984Icar...58..109F,2007prpl.conf..669L} or damping from tides raised by the star \citep{2013ApJ...774...52L}. Therefore, we caution against strict interpretations of observed exoplanet masses and architectures (or ensembles of these architectures) when it is likely that the systems have evolved substantially since their natal formation. 
We infer from the $M_\mathrm{e}/M_\mathrm{d}$ ratio that the system underwent significant orbital period changes after a migration formation, or formed in situ. %%
We also note that very high, iron-like densities are disfavored, suggesting that large amounts of collisional stripping due to high velocity giant impacts likely did not occur \citep{2010ApJ...712L..73M,2014NatGe...7..564A}.

Since the planets orbit very close to their host star, we consider the effects of tidal dissipation on the observed orbital period ratios. It is possible that tides on planets in or near a MMR causes their proximity to orbital resonance to change (generally spreading planets apart away from resonance) %($\alpha = a_2/a_1$, where $a_\mathrm{j}$ is the semi-major axis of planet j) 
over Gyr timescales \citep{2011CeMDA.111...83P,2013ApJ...774...52L}. Following \citet[][henceforth P11]{2011CeMDA.111...83P}, we define $\delta_j$ as the distance from orbital resonance by
\begin{equation}
\delta=\frac{n_{j} }{ n_{j+1}}-\frac{ (k+1) }{ k},
\end{equation}
where $n_j$ is the $j^{th}$ planet's mean motion and $k$ is the degree of the near first order resonance between planets $j$ and $j+1$.
P11 equation (40) gives the relation between the change in $\delta_j$ as a function of time and orbital parameters of the system. To determine analytically the amount tides would move planets away from exact resonance as a function of time (equation (42)), P11 integrates equation (40) from $t'=0$ to $t'=t$ and assumes $\delta_{j,t=0} = 0$, i.e., the system begins in exact MMR. If, however, we integrate from $t'=0$ to $t'=11$ Gyr (the age of Kepler-444), and we know $\delta_{j,t=11 Gyr}$ based on the observed system, we may solve for $\delta_{j,t=0}$ as a function of $Q/k_2$, the ratio of the tidal $Q$ factor and the love number. This factor enters via the tidal circularization time 
\begin{equation}
t_{c,j}=\frac{4}{63}\frac{M_j a_j^{13/2}}{(GM_\star^3)^{1/2}R_j^5}\frac{3Q}{2k_2},
\end{equation}
for the $j^{th}$ planet \citep{1966Icar....5..375G,1996ApJ...470.1187R}. We solve for the total change in distance from resonance since the planets' formation $\Delta_j=\delta_{j,t=11Gyr}-\delta_{j,t=0}$. For the inner pair of planets (b and c, $k=4$), we find that $\Delta_1\approx7\times10^{-4}-7\times10^{-7}$ for values of $Q/k_2$ ranging from 1-1000, using the approximation that $(Q/k_2)_b\approx(Q/k_2)_c$ which is reasonable given their similar size and proximity in the system. In the solar system, the rocky planets and large, rocky moons have $ 10\lesssim Q/k_2\lesssim500$ \citep{1966Icar....5..375G}. Since the observed $\delta_{1,t=11Gyr}=1.27\times10^{-2}$, we see that tidal dissipation was insufficient to have moved the innermost pair a significant distance from its current period ratio and rules out tidal dissipation breaking a natal MMR. %We do not consider 3-body interactions because the tidal effects are very small and the planets not tightly coupled due to not lying precisely at commensurabilites. 
These findings are confirmed by long-term numerical N-body integrations, following \cite{2016AJ....152..105M}. 

The other pairs of planets have longer periods, and in the case of c and d, are further from resonance. They are thus generally less affected by tides. However, the period ratio of planets d and e are very close to resonance (Table~\ref{table:prats}) so even a small amount of dissipation may significantly impact their $\delta_3$. Following \citet{2013ApJ...774...52L}, we can set a limit on the tidal $Q/k_2$ factor for the innermost planet by using their equation (18) with the observed system age and planet parameters. We find $(Q/k_2)_d\gtrsim12$. This limit is very near solar system values for rocky bodies, and possibly hints that the pair started in an exact MMR and was driven apart via this mechanism. This suggests that disk migration may have driven this pair of planets together, but the lack of tidally-broken commensurabilities among the other planets suggests the migration was not smooth or there were significant external perturbations after the disk dissipated.

\acknowledgements
We thank an anonymous referee for insightful comments which considerably added to the quality of this manuscript. This research is supported by Grant NNX14AB87G issued through NASA's \Kepler Participating Science Program.

%\bibliographystyle{apj}
%\bibliography{references}

\clearpage

\begin{table}
\caption{Kepler-444 Planet Periods and Period Ratios}
\centering
\begin{tabular}{l  |  C{0.25cm} C{0.25cm} C{0.25cm}   C{0.25cm} C{0.25cm} C{0.25cm}   C{0.25cm} C{0.25cm} C{0.25cm}  C{0.25cm} C{0.25cm} C{0.25cm}   C{0.25cm} C{0.25cm} C{0.25cm}}
\hline
&  \multicolumn{3}{c}{Planet b} &  \multicolumn{3}{c}{Planet c} &  \multicolumn{3}{c}{Planet d} &  \multicolumn{3}{c}{Planet e} & \multicolumn{3}{c}{Planet f} \\
\hline
Period (d) & \multicolumn{3}{c}{3.600105} &\multicolumn{3}{c}{4.545876} & \multicolumn{3}{c}{6.189437} &\multicolumn{3}{c}{7.743467} &\multicolumn{3}{c}{9.740501} \\
Period Ratio & && \multicolumn{2}{c}{1.262707} && \multicolumn{2}{c}{1.361550}&& \multicolumn{2}{c}{1.251078}&& \multicolumn{2}{c}{1.257899}&& \\
TTV Period (d)\tablenotemark{a} & && \multicolumn{2}{c}{89.5} && \multicolumn{2}{c}{73.1}&& \multicolumn{2}{c}{1780.3}&& \multicolumn{2}{c}{308.5}&& \\
&&&&&&&&&&&&&&&\\
\end{tabular}
\tablenotetext{1}{TTV super-period timescales calculated analytically based on the planet pair's distance from MMR \citep[see, e.g.,][]{2012ApJ...761..122L}.}
\label{table:prats}
\end{table}

\begin{table}
\centering
\caption{Photodynamic DEMCMC Posterior Median Values and 68.3\% (1-$\sigma$ equivalent) uncertainties.}
\footnotesize
\begin{tabular} { l | c | c | c | c | c }
\multicolumn{6}{l}{ \textbf{ Planet Parameters}\tablenotemark{a}}\\
\hline
& Planet b & Planet c & Planet d & Planet e & Planet f \\
\hline
$P$ (days) &$ 3.600105^{+0.000031}_{-0.000037} $		&$ 4.545876^{+0.000030}_{-0.000031} $		&$ 6.189437^{+0.000053}_{-0.000037} $ 	&$ 7.743467^{+0.000060}_{-0.00010} $ 	&	$ 9.740501^{+0.000078}_{-0.000026} $\\	
$T_0$ (days) &	$ 815.08383^{+0.00052}_{-0.00055} $	& $ 819.13903^{+0.00042}_{-0.00044} $		&$ 816.70059^{+0.00072}_{-0.00072} $ 	&$ 819.21772^{+0.00087}_{-0.00083} $ 	&	$ 817.89759^{+0.00038}_{-0.00032} $  \\
$\sqrt{e} \cos \omega$ &	$ -0.03^{+0.14}_{-0.10} $ 		&$ 0.01^{+0.12}_{-0.13} $					&$ 0.098^{+0.065}_{-0.12} $  			& $ -0.035^{+0.12}_{-0.090} $ 			&$ -0.059^{+0.12}_{-0.078} $ 	\\
$\sqrt{e} \sin \omega$ &$ 0.048^{+0.099}_{-0.15} $		&$ -0.02^{+0.13}_{-0.11} $				& $ -0.014^{+0.10}_{-0.091} $ 			&$ 0.038^{+0.074}_{-0.11} $ 			&	$ 0.052^{+0.075}_{-0.12} $ \\
$i$ ($^\circ$) &$ 92.00^{+0.26}_{-0.30} $				&$ 92.79^{+0.12}_{-0.11} $ 				&$ 91.95^{+0.11}_{-0.10} $ 			&	$ 90.62^{+0.27}_{-0.35} $			&$ 92.087^{+0.058}_{-0.054} $	\\
$\Omega$ ($^\circ$) & 0 (fixed)						&0 (fixed)								&	0 (fixed)						&0 (fixed)							&0 (fixed)	\\
$M_p / M_\star$ ($\times10^{-7}$)&	$2.3^{+1.6}_{-1.6} $  &$4.5^{+3.5}_{-3.2} $					&$1.45^{+2.6}_{-0.81} $				&$1.34^{+2.35}_{-0.74} $				& $4.5^{+12}_{-3.5} $	\\
$R_p / R_\star$ ($\times10^{-3}$) &$ 4.967^{+0.070}_{-0.067} $	&$ 6.380^{+0.090}_{-0.087} $		&$ 6.613^{+0.079}_{-0.077} $ 			&$ 6.799^{+0.078}_{-0.076} $ 			&$ 9.39^{+0.13}_{-0.12} $	\\
\multicolumn{6}{l}{}
\end{tabular}
\begin{tabular} { l | c }
\multicolumn{2}{l}{\textbf{Stellar Parameters}}\\
\hline
$M_\star$ ($M_\odot$)\tablenotemark{b}  &$0.758\, (\pm0.043)$\\
$R_\star$ ($R_\odot$) &$0.749^{+0.014}_{-0.013}$ \\
$c_1$ &$ 0.45^{+0.13}_{-0.14}$	\\
$c_2$ &$0.32^{+0.20}_{-0.19}$	\\
$dilute$ &0.0394 (fixed)	\\
\multicolumn{2}{l}{}
\end{tabular}
\begin{tabular} { l | c | c cc c | c cc c}
\multicolumn{10}{l}{\textbf{Planet Mass Posteriors Convolved with Stellar Uncertainties}\tablenotemark{b}}\\
\hline
Planet & Radius  & Median Mass & 68.3\% CI &  95\% CI & 99\% CI & Density &  68.3\% CI &  95\% CI & 99\% CI  \\
&($R_\mathrm{\oplus}$) & ($M_\mathrm{\oplus}$) & && & (g cm$^{-3}$) && & \\
\hline 
b &$ 0.406 ^{+     0.013 } _{-    0.013 } $	&			&  $< 0.079$ 				& $< 0.11$			&$< 0.13$				& 			&$< 6.6$ 				&$<9.1 $				&  $<9.4 $	 \\
c & $0.521 ^{+     0.017 } _{-     0.016} $	& 			& $<0.16 $ 				&$< 0.24$ 			& $< 0.27$			&			& $<  6.2$				& $<9.1 $				&   $<9.6 $     \\
d &$0.540 ^{+     0.017 } _{-     0.016 }$	&$ 0.036$		&$[     0.016 ,      0.10 ]$		&$ [    0.0092 ,      0.20 ]$	&$[    0.0070 ,      0.27 ]$ 	& $1.27 $		&$[       0.56 ,        3.5 ]$ 	& $[       0.32 ,        7.2]$	&  $[       0.25 ,        9.2 ]$     \\
e & $0.555 ^{+     0.018 } _{-     0.016 }$	&$0.034 $		&$[     0.015 ,      0.093] $  	&$[    0.0087 ,      0.19 ]$	&$[    0.0065 ,      0.25 ]$	& $ 1.08 $		&$[       0.48 ,        3.0 ]$ 	& $[       0.28 ,        6.1 ] $ 	& $[       0.21 ,        8.0 ]$	\\
f & $0.767 ^{+     0.025 } _{-     0.024 }$	& 			& $< 0.22$				& $< 0.71$			&$< 0.94$				&			&$< 2.6$ 				&$< 8.8$ 				& $<11 $	\\
\end{tabular}
\tablenotetext{1}{Valid at $T_\mathrm{epoch} = 815$ (BJD - 2454900 days)}. 
\tablenotetext{2}{$M_\star$ is held fixed at $0.758M_\odot$ in the DEMCMC, but the posteriors are convolved with the uncertainties on stellar mass ($0.758\pm0.043M_\odot$) from \cite{2015ApJ...799..170C} when determining uncertainties in physical units in the bottom panel.}
\label{table:allparams}
\end{table}

\end{document}